\documentclass[12pt,a4paper]{article}
\usepackage[left = 2cm,right = 2cm]{geometry}
\usepackage{amssymb}
\usepackage{graphicx,amsmath,bm}
\usepackage{amsmath}
\usepackage{braket}
\usepackage{url}
\usepackage{subfigure}
\usepackage{graphicx}
\usepackage{booktabs}
\usepackage{multirow}
\usepackage{hyperref}
\usepackage{caption}

\begin{document}
\title{Neutron-Proton Interaction\\ Modeled using Morse Function:\\ Constructing Inverse Potentials Using Variational Monte-Carlo and Phase Function Method}
\author{Anil Khachi, Lalit Kumar and O.S.K.S. Sastri$^*$\\Department of Physics and Astronomical Sciences, \\Central University of Himachal Pradesh, H.P., Bharat(India)}
\maketitle
\begin{abstract}
Understanding neutron-proton(np) interaction has been one of the most studied problems. One way to construct model interaction has been using inversion potentials obtained from experimental scattering phase shifts(\textit{SPS}). Here, we show that, inverse potentials corresponding to \textit{SPS} for various $\ell$-channels of np interaction can be obtained using variational Monte-Carlo(\textit{VMC}) technique in tandem with phase function method(\textit{PFM}) by modeling np-interaction as a Morse function. The S-channel \textit{SPS} for $^3S_1$ and $^1S_0$ have been obtained, with a mean percentage error with respect to experimental multiple energy analysis data for lab energies upto 1050 MeV, to less than 5$\%$. Similarly, inverse potential for $^1S_0$ pp interaction, with Coulomb term modeled as proportional to erf(), has also been obtained to match experimental values to less than 4$\%$. Non-local and spin-orbit terms are included to obtain inverse potentials for P and D channels and results match with available data to a good extent. 
\end{abstract}
\section{Introduction}
\label{intro}
The neutron-proton interaction has been first modeled by Yukawa\cite{1}. This was followed by various single and multi-particle exchange models and QCD based models as detailed in these reviews\cite{2,3}. Currently, the Nijm\cite{4} and Argonne\cite{5} potentials are the ones which give rise to best quantitative results for explaining the experimental scattering phase shifts. Unfortunately, all these potentials have different mathematical representations originating from completely varied physical considerations. Yet all of them lead to correct validation of experimental data. The search for a simple theoretical potential that could model the nucleon-nucleon interactions is still eluding the physicists. Interestingly, many simple phenomological forms such as Square well, Malfiet-Tjohn\cite{6}, Hulthen\cite{7}, have also been utilised for studying the deuteron. Recently, a molecular Manning-Rosen\cite{8} potential has been proposed. Some of these works involve partial wave analysis by proposing symmetric partner potentials, based on super symmetric (SUSY) quantum mechanics, as the centrifugal barriers for higher $\ell$-values. It is more or less well established now that the strong force is a result of quark interactions within hadrons and mesons, and the nucleons themselves experience an attractive force which is of secondary nature akin to Vander Wall interaction within two neutral atoms in a molecule. Based on this premise and understanding of the characteristics of n-p interaction which has a repulsive core at short inter-nucleon  distance ($< 0.5 fm$), followed by an attractive nature between $0.5 - 2$ fm and an exponentially decaying tail as the nucleons go far apart, we have considered in our recent work\cite{9, 10}, the most successfull molecular Morse potential to describe phenomologically the np interaction.\\
Alternatively, inverse potentials resulting directly from experimental observations by J-matrix method\cite{11} and Marchenko equation\cite{12, 13} have also found some success in understanding the interaction involved between the nucleons. In this paper, we propose a purely computational approach to obtaining inverse potentials responsible for scattering phase-shifts (SPS) in nucleon-nucleon interactions. While the SPS are determined using phase function method (PFM), the model parameters are optimized to obtain best match with experimental mean energy analysis data (\textit{MEAD}) using a slightly modified variational Monte-Carlo(VMC) technique\cite{14}.    \\
\section{Methodology:}
The Morse function considered as the model of interaction for the weakly bound deuteron is given by 
\begin{equation}
V_1(r) = V_0\left(e^{-2(r-r_m)/a_m)}-2e^{-(r-r_m)/a_m}\right)
\label{eq1}
\end{equation} 
and it belongs to the class of potentials that are shape-invariant\cite{15}. Here, the model parameters $V_0, r_m$ and $a_m $ reflect strength of interaction, equilibrium distance at which the maximum attraction is felt and shape of potential respectively. 
\subsection{Phase Function Method:}  
The first order non-homogeneous differential equation for \textit{SPS} \cite{16,17,18} is given by 
\begin{equation}
\delta_{\ell}'(k,r)=-\frac{V(r)}{k}\bigg[cos(\delta_\ell(r))\hat{j}_{\ell}(kr)-sin(\delta_\ell(r))\hat{\eta}_{\ell}(kr)\bigg]^2
\label{PFMeqn}
\end{equation}
This is numerically solved using Runge-Kutta 5th order method. For $\ell = 0$, the Ricatti-Bessel and Riccati-Neumann functions $\hat{j_0}$ and $\hat{\eta_0}$ get simplified as $sin(kr)$ and $-cos(kr)$, respectively. 

The greatest advantage of this method is that, the phase-shifts are directly expressed in terms of the potential and have no relation to the wavefunction. This has been utilised in this paper to obtain inverse potentials in an innovative way by implementing a modified VMC in tandem with PFM. The technique optimizes the model parameters of the potentials to obtain the best match with respect to the experimental values.

\subsection{Optimization of Morse function model parameters using VMC:}
Typically, VMC is utilised for obtaining the ground state energy for a given potential. The method starts with a trial wavefunction, which is varied at a random location by a random amount in the Monte-Carlo sense. Then, the energy is determined using the newly obtained wavefunction and variational priniciple is applied. This process is done iteratively till one converges to the ground state. Here, we consider to vary the potential instead of wavefunction and minimise variance with respect to experimental data, as follows:
\begin{description}
\item [Initialisation step:] To begin the optimisation procedure, Morse parameters $V_0$, $r_m$ and $a_m$ are given some initial values. The phase equation is integrated using RK-5 method for different values of k, a function of lab energies E, to obtain the simulated SPS, say \textit{$\delta^{sim}_k$}. The mean percentage error (\textit{MPE}) has been determined with respect to the experimental mean energy analysis data of Arndt\cite{19}, say \textit{$\delta^{exp}_k$}, as
\begin{equation}
MPE = \frac{1}{N}\sum_{i=1}^N \frac{|\delta^{exp}_k - \delta^{sim}_k|}{|\delta^{exp}_k|} * 100
\end{equation}
This is named as $MPE_{old}$ and is also assigned to $MPE_{min}$.
\item[Monte-Carlo step:] A random number $r$, generated in an interval [-I, I], is added to one of the parameters, say $V_{0new} = V_0 + r$.
\item[PFM step:] Again, the phase equation is integrated with new set of parameters $V_{0new}$, $r_m$ and $a_m$ to obtain new set of simulated SPS, say \textit{$\delta^{sim-new}_k$}, using which $MPE_{new}$ is determined.
\item[Variational step:] If $MPE_{new} < MPE_{old}$, then $V_0 = V_{0new}$, $MPE_{min} = MPE_{new}$,  else old values are retained. 
\end{description}
The final three steps are repeated for each of the parameters to complete one iteration. The size of interval is reduced after a certain number of iterations, if there is no significant reduction in $MPE_{min}$. The process is completed when $MPE_{min}$ does not change any further, that is, convergence is reached. 
 
\section{Results:}
\subsection{np scattering phase shifts for $^3S_1$:}
The analytical solution of time independent Schrodinger equation for Morse potential\cite{20}, in Eq. \ref{eq1}, is given by
\begin{equation}
E_n = -\frac{\hbar^2}{2\mu a_m^2}(p-n-1/2)^2
\label{eigenvals}
\end{equation}
where
\begin{equation}
p^2 = \frac{2\mu V_0 a_m^2}{\hbar^2}
\end{equation}
The energy expression is only consisting of $V_0$ and $a_m$. Using the binding energy of deuteron $E_D = -2.224589$ MeV for the ground state, one can impose a constraint on $V_0$ for a given value of $a_m$. So, we need to vary only two parameters, $a_m$ and $r_m$ in VMC while obtaining the SPS using PFM. The \textit{MPE} has been reduced to less than $4\%$ for both \textit{MEAD} of 350 and 1050 MeV for parameters shown in Table \ref{table1}. This procedure is akin to obtaining the inversion potential that results in best \textit{SPS}, while retaining the binding energy of deuteron. \\ 
A comparative analysis parameter that represents relative mean square error, is defined as 
\begin{equation}
c^2 = \frac{1}{N} \sum_{i=1}^N \frac{\left(\delta_i^{expt} -\delta_i^{sim}\right)^2}{\delta_i^{expt}}
\end{equation}
This parameter has been computed by comparing simulated SPS with \textit{MEAD} of Arndt \cite{19} and found to be 0.24(0.16) for data upto 350(1050) MeV.
Typically, a value less than 1 is considered to be a good match between the two sets of data.
\begin{figure*}[htp]
\centering
\includegraphics[scale=0.5]{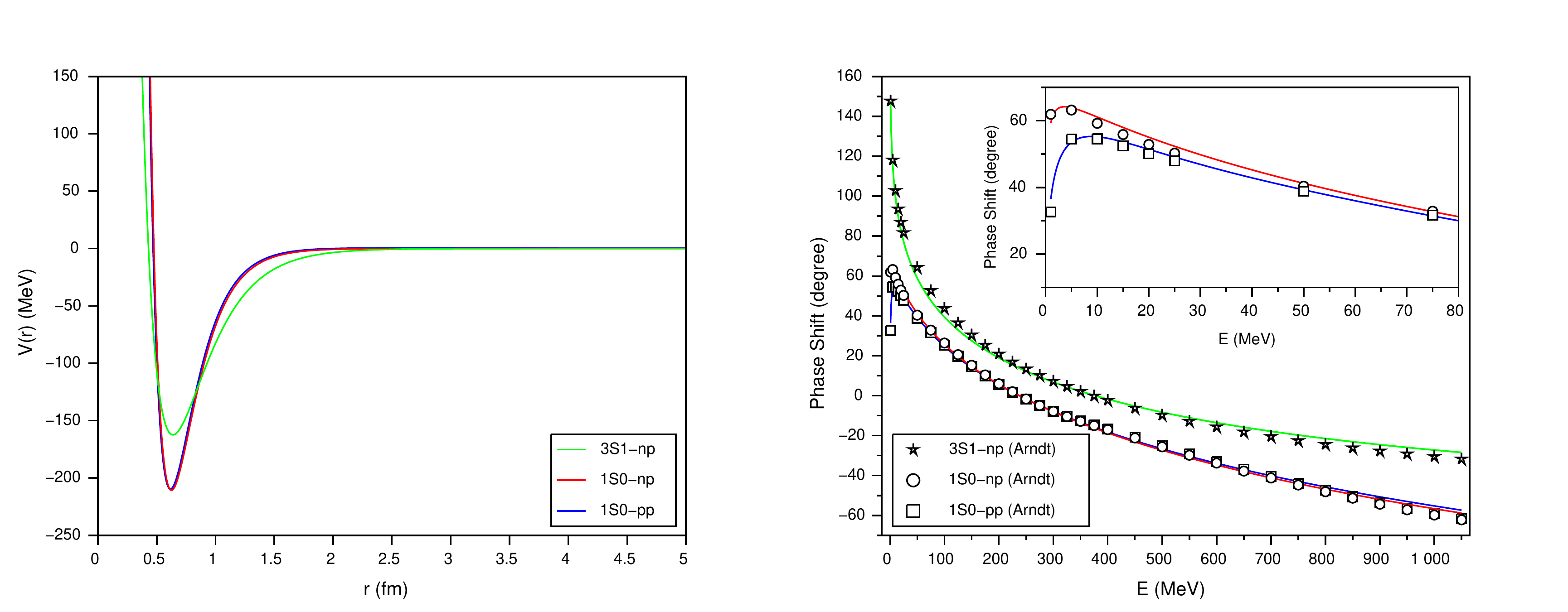} 
\caption{(a) Potentials for S-states of np and pp interactions based on Morse function. (b) Corresponding simulated scattering phase-shifts using \textit{PFM}, along with respective experimental data from Arndt \cite{19}. Inset shows detail for $^1S_0$ np $\&$ pp \textit{SPS} for lab energies less than 100 MeV.}
\label{fig1a}
\end{figure*}

\subsection{np scattering potential for $^1S_0$:}
The singlet S-state, for anti-parallel spins given by $^1S_0$, is not a bound state. Starting with $^3S_1$ potential parameters, \textit{SPS} of $^1S_0$ are determined using \textit{PFM}. Then, \textit{MPE} is calculated w.r.t \textit{MEAD} of $^1S_0$. Now, \textit{VMC} and \textit{PFM} are iterated till \textit{MPE} converges to a constant value. The resultant parameters of inverse potential, see Fig \ref{fig1a}(a), are given in Table \ref{table1}. The obtained  \textit{SPS} for all energies upto 1050 MeV are shown in Fig. \ref{fig1a}(b). The \textit{MPE} and c$^2$ for data upto 350 (1050) MeV are found to be 2.8$\%$ (3.6$\%$) and 0.03 (0.05) respectively. \\ 
 
\subsection{pp scattering potential of $^1S_0$:}
To describe pp interaction, an extra Coulomb term\cite{21} is added to np interaction potential, as 
\begin{equation}
V_{Coulomb} = \frac{e^2}{r}erf(\beta r)
\end{equation}
Where $\beta$ is a parameter to be adjusted for obtaining correct experimental \textit{SPS}. 
Starting with parameters obtained for $^1S_0$ \textit{SPS}, $\beta$ is also varied in Monte-Carlo sense to minimize \textit{MPE} w.r.t. experimental \textit{SPS} due to pp interaction. For $\beta = 0.75 fm^{-1}$, minimum \textit{MPE} value of 2.8$\%$(3.8$\%$) and $c^2$ value of 0.03(0.07) for data up to 350(1050) MeV are obtained. The potential and \textit{SPS} plot are shown in Fig. \ref{fig1a}. 

\subsection{Inverse potentials for P and D states:}
In order to obtain \textit{SPS} shifts for higher $\ell$ values, one needs to bring in a non-local potential form that contains $\ell$ dependence. This is proposed to take care of all possible spin-spin, isospin-isospin, and spin-isospin dependencies. Here, in this paper, we assume non-local potential to have a similar form as that of local potential and utilized Aldrich-Green approximation \cite{22} as 
\begin{equation}
V_2(r) = \frac{\hbar^2}{2\mu}\ell(\ell+1)\bigg[\frac{e^{-r/a_m}}{a_m^2}\frac{1}{(1-e^{-r/a_m})^2}\bigg]
\label{eq2}
\end{equation}
There are no new parameters incorporated into it. Since we are dealing with partial wave analysis of individual channels, the tensor part, and its components, resulting from various spin and isospin dependencies, have been assumed to be internal to central part of the potential. While this suffices for singlet P and D channels, one has to include spin-orbit interaction for triplet states, given by
\begin{equation}
V_{LS}(r)={\left(\frac{r_0}{\hbar}\right)}^2\frac{1}{r}\frac{2V_0}{a_m}\bigg[e^{-(r-r_m)/a_m}-e^{-2(r-r_m)/a_m}\bigg] (\vec{L}.\vec{S})
\label{eq3}
\end{equation}
Here, $\vec{L}.\vec{S} = \frac{\hbar^2}{2}[J(J+1)-L(L+1) - S(S+1)]$. The proportionality constant $r_0^2$ takes care of dimensional analysis and also adds an extra parameter for optimization. \\
The phase function equation for $\ell=1$, i.e. P-partial wave, is of the form
\begin{equation}
\delta_1'(k,r)=-\frac{V(r)}{k}\bigg[cos(\delta_1(r))\big(\frac{sin kr}{kr}-coskr\big)+sin(\delta_1(r))\big( \frac{cos kr}{kr}+sinkr \big)\bigg]^2
\end{equation}
and that for $\ell=2$, i.e. D-partial wave, is of the form
\begin{equation}
\delta_2'(k,r) = -\frac{V(r)}{k}[cos(\delta_2(r))f(kr) - sin(\delta_2(r))g(kr)]^2
\end{equation}
where f(kr) is 
\begin{equation}
f(kr) = \left(\frac{3}{(kr)^2}  + 1\right)sinkr-\frac{3}{kr}
\end{equation}
and g(kr) is 
\begin{equation}
g(kr) = \left(\frac{-3}{(kr)^2} + 1\right)coskr- \frac{3}{kr}sinkr
\end{equation}

The procedure of \textit{VMC} in tandem with \textit{PFM} equations as given above has been applied to all states of $\ell$ = 1 $\& $ 2 respectively. One has to include the non-local potential for singlet states and add the spin-orbit potential as well for the triplet states, while determining the SPS. The obtained inversion potential parameters, along with \textit{MPE} and $c^2$ values, are shown in Table \ref{table1} .
\begin{table*}
\centering
\caption{Optimized inversion potential parameters obtained for S, P and D channels for np interaction and singlet S channel's pp interaction are presented. The respective mean percentage error (\textit{MPE}) and comparative analysis parameter ($c^2$) defined as the relative mean squared error w.r.t. \textit{MEAD} of Arndt\cite{19} are shown for all partial waves.}
\label{table1}
\scalebox{0.9}{
\begin{tabular}{|c|cccccc|cccccc|} 
\hline
\multirow{2}{*}{ \textbf{States} }          & \multicolumn{6}{c|}{\textbf{350 MeV} }                                                                                                                                                                                     & \multicolumn{6}{c|}{\textbf{1050 MeV} }                                                                                                                                                                                     \\ 
\cline{2-13}
                                            & \multicolumn{1}{c|}{\textbf{$V_{0}$} } & \multicolumn{1}{c|}{\textbf{$r_{m}$} } & \multicolumn{1}{c|}{\textbf{$a_{m}$} } & \multicolumn{1}{c|}{\textbf{$\beta$} } & \multicolumn{1}{c|}{\textbf{$c^2$} } & \textbf{$MPE$}  & \multicolumn{1}{c|}{\textbf{$V_{0}$} } & \multicolumn{1}{c|}{\textbf{$r_{m}$} } & \multicolumn{1}{c|}{\textbf{$a_{m}$} } & \multicolumn{1}{c|}{\textbf{$\beta$} } & \multicolumn{1}{c|}{\textbf{$c^2$} } & \textbf{$MPE$}   \\ 
\hline
\textbf{$^{3}S_{1}$}                        & 256.5204 & 0.5628                                  & 0.2217                                   & --                                     & 0.24                              & 3.6            & 213.935                                & 0.582                                                                   & 0.245                                   & --                                      & 0.16 & 4.6             \\
\textbf{$^{1}S_{0}-np$}                     & 216.385                                & 0.623                               & 0.2125                                 & --                                     & 0.03                                 & 2.8             & 216.385                               & 0.623                                 & 0.2125                                 & --                                      & 0.05                                 & 3.6              \\
\textbf{$^{1}S_{0}-pp$}                     & 209.805                                & 0.617                                  & 0.214                                  & 0.750                                  & 0.03                                 & 2.8               & 209.805                                & 0.617                                  & 0.214                                  & 0.750                                  & 0.07                                 & 3.8              \\ 
\hline
\begin{tabular}[c]{@{}c@{}}\\ \end{tabular} & \multicolumn{1}{c|}{\textbf{$V_{0}$} } & \multicolumn{1}{c|}{\textbf{$r_{m}$} } & \multicolumn{1}{c|}{\textbf{$a_{m}$} } & \multicolumn{1}{c|}{\textbf{$r_{0}$} } & \multicolumn{1}{c|}{\textbf{$c^2$} } & \textbf{$MPE$}  & \multicolumn{1}{c|}{\textbf{$V_{0}$} } & \multicolumn{1}{c|}{\textbf{$r_{m}$} } & \multicolumn{1}{c|}{\textbf{$a_{m}$} } & \multicolumn{1}{c|}{\textbf{$r_{0}$} } & \multicolumn{1}{c|}{\textbf{$c^2$} } & \textbf{$MPE$}   \\ 
\hline
\textbf{$^{1}P_{1}$}                        & 5.162                                  & 1.345                                  & 0.404                                  &--                                      & 0.15                                 & 16.1            & 3.398                                  & 1.240                                  & 0.418                                  &--                                      & 0.29                                 & 13.7             \\
\textbf{$^{3}P_{0}$}                        & 77.295                                 & 0.917                                  & 0.397                                  & 0.256                                  & 0.07                                 & 7.9             & 95.104                                 & 0.977                                  & 0.289                                  & 0.075                                  & 0.26                                 & 13.6             \\
\textbf{$^{3}P_{1}$}                        & 7.385                                  & 0.711                                  & 0.832                                  & 2.292                                  & 0.03                                 & 2.5             & 0.888                                  & 1.233                                  & 0.363                                  & 0.786                                  & 0.18                                 & 12.7             \\
\textbf{$^{3}P_{2}$}                        & 548.605                                & 0.296                                  & 0.235                                  & 0.021                                  & 0.03                                 & 6.1             & 456.125                                & 0.484                                  & 0.149                                  & 0.031                                  & 0.45                                 & 22.8             \\ 
\hline
\textbf{$^{1}D_{2}$}                        & 501.315                                & 0.450                                  & 0.304                                  & --                                      & 0.10                                 & 26.4            & 288.927                                & 0.695                                  & 0.205                                  & --                                     & 0.19                                 & 24.6             \\
\textbf{$^{3}D_{1}$}                        & 175.130                                & 0.083                                  & 0.651                                  & 0.205                                  & 0.004                                & 3.7             & 175.185                                & 0.035                                  & 0.599                                  & 0.301                                  & 0.02                                 & 3.5              \\
\textbf{$^{3}D_{2}$}                        & 262.058                                & -0.325                                 & 0.735                                  & 1.400                                  & 0.29                                 & 17.1            & 262.693                                & -0.297                                 & 0.705                                  & 1.442                                  & 0.22                                 & 13.2             \\
\textbf{$^{3}D_{3}$}                        & 486.205                                & 0.462                                  & 0.282                                  & 0.131                                  & 0.01                                 & 11.2            & 486.482                                & 0.475                                  & 0.239                                  & 0.093                                  & 0.06                                 & 23.8             \\
\bottomrule
\end{tabular}
}
\end{table*}
It is observed that triplet P-states are obtained to an accuracy of less than $8\%$ for data up to 350 MeV and singlet is having an \textit{MPE} of about 16$\%$.

\begin{figure*}[htp]
\centering
{\includegraphics[scale=0.5]{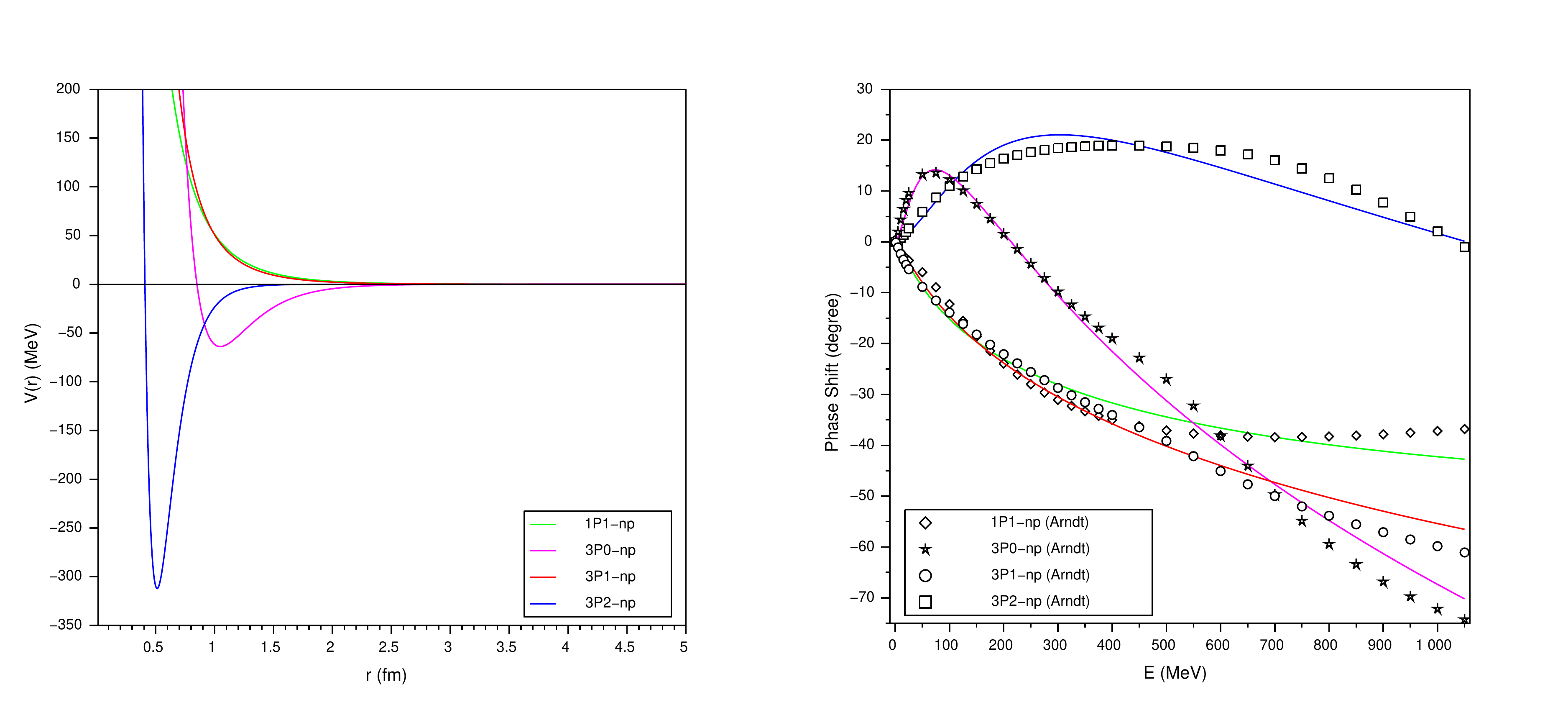}  }
\caption{(a) Inversion potentials for singlet $^1P_1$ and triplet $^3P_{0,1,2}$ states.\\(b) Corresponding computed scattering phase-shifts along with their respective experimental data from Arndt\cite{19}}
\label{fig2}
\quad
{\includegraphics[scale=0.5]{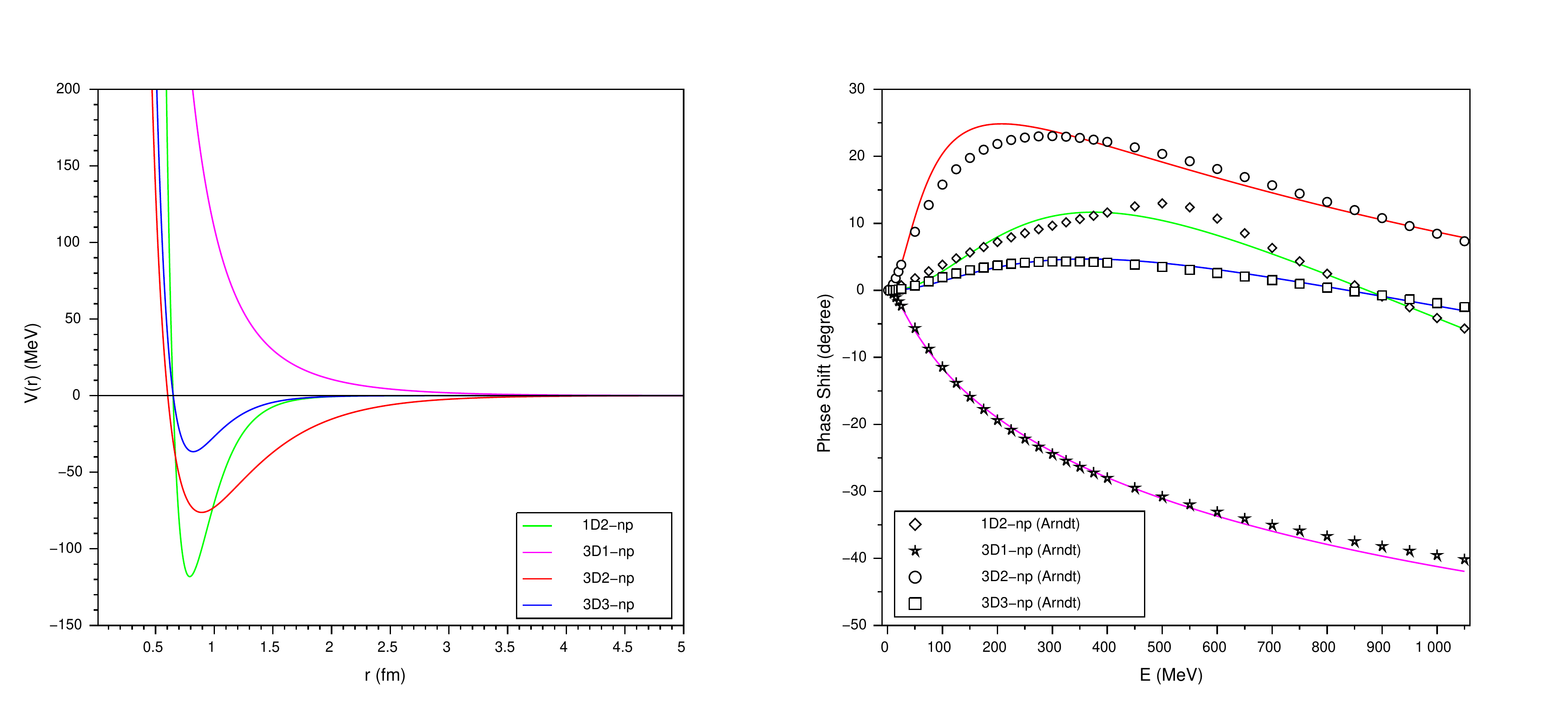}  }
\caption{(a) Inversion potentials for singlet $^1D_2$ and triplet $^3D_{1,2,3}$ states.\\ (b) Corresponding computed scattering phase-shifts along with their respective experimental data from Arndt\cite{19}}
\label{fig3}
\end{figure*}

Similarly, results obtained for triplet D-states are slightly better than that of singlet D-state. This clearly shows that having an extra parameter due to the spin-orbit term has given more options for \textit{VMC} to converge better as compared to singlet terms which did not have any extra parameter to adjust. The obtained \textit{SPS} for various P and D states have been shown in Fig. \ref{fig2}(b) and Fig. \ref{fig3}(b), respectively. The matching trends w.r.t \textit{MEAD} of Arndt\cite{19} indicate the merit of our procedure in obtaining the inverse potentials. Especially, the trends in various curves reveal a good match for data up to 350 MeV. The potentials do show resemblance to what has been obtained by Nijemen group\cite{4}. 

\section{Discussion}
\subsection{Effectiveness of Morse function:}
\begin{enumerate}
\item We have undertaken a comparative study\cite{9} by considering phenomenological two term Malfiet-Tjon \cite{6} and Manning-Rosen \cite{8}, potentials and found that $^3S_1$ \textit{SPS} obtained with these saturate to a constant value above a certain lab energy.

\item An attempt has been made to include tensor interaction due to $^3D_1$ state, modeled as proportional to ($1/r^2$)\cite{23}, so that the actual central part of potential could be obtained. But, it turned out that $^3S_1$ \textit{SPS} are better represented with Morse function. Hence, tensor interaction has been assumed to be an inherent part of it. We also observe that best convergence for \textit{SPS} has been obtained for $^3D_1$ state as compared to all other partial D-wave channels.
\end{enumerate}

\subsection{Simple phenomenology for pp interaction:}
\begin{enumerate}
\item We have shown that $^1S_0$ np \textit{SPS} can be obtained using same local interaction potential with albeit slightly different model parameters.
\item Also, \textit{SPS} for $^1S_0$ pp interaction have been successfully matched with \textit{MEAD} of Arndt, by modelling the Coulomb interaction as proportional to an $erf()$ function with only one extra parameter.
\item It is simplicity of Morse form that has been able to achieve this convergence. One of the important characteristics of a good model is parsimony.
\end{enumerate}
\subsection{Non-local interactions and inversion potentials:}
The non-local interaction to model spin and iso-spin dependencies based on Aldrich-Green approximation \cite{22} of centrifugal term, does provide impetus to the inversion technique, as shown in our results. But, there are other considerations that might further improve convergence. For example, the methodology of super-symmetric quantum physics(\textit{SUSY}) to construct partner super-potentials\cite{7} for $\ell \neq 0$ scattering states from its hypergeometric type ground state wave-functions might be a better choice and is being attempted. 
\subsection{Inversion potentials for P and D channels:}
While performing \textit{VMC}, initial values play an important role in the convergence process. Of course, same starting points could lead to a slightly different set of parameters on convergence, for a chosen number of decimal places, as is expected in a \textit{VMC} procedure. Also, completely different starting points might also end up giving similar convergences as well, in which case, it is important to look at the corresponding potential plots to get a better understanding. 
Here are a few observations based on our multiple runs with regard to the construction of inversion potentials for P and D channels.:
\begin{enumerate}
\item In case of $^3D_3$, it was found that by starting with a completely different set of starting values, \textit{VMC} ended up with $V_0$ = 269.743, $r_m$ = 0.671, $a_m$ = 0.286 and $r_0$ = 0.1995 yielding a better convergence for \textit{MPE} as 8$\%$ as compared to current values quoted in Table \ref{table1} which have resulted in \textit{MPE} of 11.2 $\%$. On plotting the potentials with both sets of parameters, it was observed that potential due to former parameters has resulted in an extremely large attractive part with a magnitude close to $10^{5}$ for r values between 0.03-0.04 fm, followed by another low attractive potential between 1-1.3 fm and finally the decaying tail part. This is certainly an aberration, as such strong attraction, is not anticipated in any physical interaction. A very similar occurrence has been observed in the case of $^3P_2$ as well.
\item Multiple runs for $^3D_2$ have shown that, as initial parameters values are slightly varied, the convergence to final parameters lie within a small range. For instance, values of [$V_0, r_m, a_m$, $r_0$] have been obtained as [272.172, -0.070, 0.657, 1.148] and [252.431, -0.234,   0.719, 1.344] with \textit{MPE} values of 17.5$\%$ and 17.3$\%$ respectively. Even though $V_0$ in Table \ref{table1} appears to be the mean of these two quoted sets, all other parameters are higher. So there are no particular correlations to guide the process of convergence.
\item In case of singlet channels, $^1P_1$ and $^1D_2$, we have included a proportionality constant $r_1$ to act as a parameter to check if the convergence improves. This resulted in very small gains. For example, the value of \textit{MPE} reduced from 16.1$\%$ to only 15.8$\%$ in case of $^1P_1$ \textit{SPS}, for data up to 350 MeV. Here, we have noted that three sets of completely different parameters [95.507, 0.642, 0.422, 2.238], [194.245, 0.366, 0.451, 2.464] and [240.561, 0.354, 0.417, 2.755] have all resulted in \textit{MPE} values which are close to 15.9$\%$. Even though the parameter sets are extremely different, their potential plots are all very closely overlapping and are hardly distinguishable. We were expecting that there would be a possibility of obtaining isospectral potentials, just as in case of Marchenko technique \cite{9}, but such a scenario does not appear to be occurring in our procedure. One might have to perform many more such trials before one can rule out completely such a possibility.
\end{enumerate}
\subsection{Application to other problems:}
Our group has applied this technique to obtain best model parameters for Woods-Saxon potential\cite{24} for determining single-particle energies of neutron and proton states in Shell model of the nucleus. Similarly, we have also applied this technique to obtain best model parameters for Morse potential\cite{14} to determine ro-vibrational frequencies of diatomic molecule HCl. It has been observed that results obtained are better than those from regression analysis performed on analytical expressions based on theoretical considerations. In fact, one can conclude that \textit{VMC} is an equivalent methodology for least-squares minimization-based optimization. Even though \textit{VMC} procedure looks very promising, one must keep in mind that it strongly depends on initial conditions. There is every possibility that optimization process could get stuck in a local minima and not lead to convergence. This is one of the reasons why a good theoretical framework could be conisdered to act as a guide to \textit{VMC} procedure.

Finally, to conclude Morse function seems to be an ideal phenomenological potential that is able to explain np interactions in the weakly bound deuteron system. Taken along with Coulomb term as proportional to an $erf()$ it is able to represent pp interaction as well. The inversion methodology based on \textit{VMC} needs to be performed for higher $\ell > 2$ channels. This could lead to reconstruction of various internal interaction potentials that in turn could give complete quantitative description of experimental data. Further, our \textit{VMC} least square error minimization procedure might prove very useful in chemical and biological systems, economics, management, and mathematics and where ever optimization of model parameters is involved.  	
\section*{Acknowledgement}
We would like to thank Prof R. C. Verma for his insightful discussions.

\end{document}